\documentclass[%
 reprint,
amsmath,amssymb,
aps,
]{revtex4-1}
\usepackage{graphicx}
\usepackage{dcolumn}
\usepackage{bm}
\usepackage{epstopdf}

\begin{document}

\preprint{APS/123-QED}

\title{Electron carriers with possible Dirac-cone-like dispersion in FeSe$_{1-x}$S$_x$ ($x$ = 0 and 0.14) single crystals triggered by structural transition}

\author{Yue Sun,}
 \email{sunyue.seu@gmail.com}
\author{Sunseng Pyon, and Tsuyoshi Tamegai}

\affiliation{%
Department of Applied Physics, The University of Tokyo, 7-3-1 Hongo, Bunkyo-ku, Tokyo 113-8656, Japan}
\date{\today}

\begin{abstract}
We report detailed study of the transport properties of  FeSe$_{1-x}$S$_x$ ($x$ = 0 and 0.14) single crystals grown by vapor transport method. 14\% S doping is found significantly suppress the structural transition from $T_s$ $\sim$ 86 K in FeSe to $\sim$ 49 K, although the superconducting transition temperature, $T_c$, is only slightly affected. A  pronounced linear magnetoresistance (MR) is observed in both FeSe and FeSe$_{0.86}$S$_{0.14}$ single crystals, which is found to be triggered by the structural transition. The linear MR and related discussion indicate the possible existence of Dirac-cone-like state, which may come from the band shift induced by ferro-orbital order. The mobility of the Dirac-cone-like band is found to decrease after S doping. Besides, the invalid Kohler's scaling of MR is found for temperature below $T_s$ in both crystals, however the re-establishment of the Kohler's scaling at temperatures below 30 K is observed in FeSe, but not in FeSe$_{0.86}$S$_{0.14}$. All these observations above support that the orbital ordering causes the band reconstruction in FeSe, and also that the orbital ordering in FeSe is suppressed by the chemical pressure from S doping.

\begin{description}
\item[PACS numbers]
\verb+74.70.Xa+, \verb+74.25.F-+, \verb+72.15.Gd+, \verb+75.47.-m+

\end{description}
\end{abstract}

\pacs{Valid PACS appear here}
\maketitle
\section{introduction}
FeSe, composed of only Fe-Se layers \cite{HsuFongChiFeSediscovery}, has the simplest crystal structure in iron-based superconductors (IBSs) and is also one of the most intriguing candidates for both searching high-temperature superconductivity and probing the superconducting mechanism. Although the initial $T_c$ in FeSe is below 10 K \cite{HsuFongChiFeSediscovery}, it can be easily increased to 37 K under pressure \cite{MedvedevNatMat} and over 40 K by intercalating space layers \cite{BurrardNatMat,SunLilingnature}. Recently, the monolayer of FeSe grown on SrTiO$_3$ is reported to show a sign of superconductivity over 100 K \cite{GeNatMatter}.

Different from the iron pnictide, in which a tetragonal to orthorhombic structural transition usually precedes or coincides with stripe-type antiferromagnetic (AFM) order in close proximity with pressure or chemical doping \cite{FernandesNatPhy}, FeSe undergoes only the structural transition at $T_s$ $\sim$ 87 K without long-range magnetic order at any temperature \cite{McQueenPRL}. Such a unique feature makes FeSe an ideal material to study the nematic order, which is often referred as the origin of structural transition and is believed to be related directly to the high-temperature superconductivity \cite{FernandesNatPhy,ChuScience,FradkinAnnu}, without the influence of magnetic order. Recent angle-resolved photoemission spectroscopy (ARPES) study on FeSe reported a splitting of the otherwise degenerate Fe 3$d_{xz}$ and 3$d_{yz}$ orbitals at the $M$ point of the Brillouin zone. Such splitting of bands is as large as 50 meV at low temperatures and can persists up to temperatures of $\sim$ 110 K above $T_s$, which indicates that the electronic nematicity is caused by the ferro-orbital ordering \cite{NakayamaPRL,ShimojimaPRB}. It is also supported by the NMR measurements that spin fluctuations only exists below $T_s$, which is against spin-driven nematicity \cite{BaekNatMater,BöhmerPRL}.

On the other hand, in the nematic state, FeSe also manifests some special transport properties like the non-linear Hall resistivity and a sudden sign reversal in the temperature dependence of the Hall coefficient, together with an emergence of large magnetoresistance (MR) \cite{LeiHechangPRBMultiband,HuynhPRB,WatsonPRL}. Besides, the mobility spectrum analysis and the three-carrier model fitting all suggest a possible electron-type carrier with ultrahigh mobility emerges below $T_s$  \cite{HuynhPRB,WatsonPRL}. However, until now the origin of the distinguished transport property and its relation with the structural transition in FeSe is still not well studied, which is crucial to the understanding of its band structure as well as the novel superconductivity. Initially, the S-doping effect in FeSe has been studied in polycryatslline samples \cite{MizuguchiJPSJ}. Recently, high-quality single crystals of S-doped FeSe have been successfully grown by vapor transport method, and it is reported that their superconducting properties are affected little by the S-doping \cite{AbdelPRB}. APRES studies found that the S doping reduces the $T_s$ possibly by the suppression of orbital ordering \cite{Watsonarxiv}. Since the S is isovalent to FeSe, it will not introduce extra charge carriers. And its nonmagnetic nature will not also disturb the scattering too much. Thus, the comparative study of transport properties in pure and S-doped FeSe is helpful to the understanding of the extraordinary transport properties and their relation with structural transtion/nematic order.

Unfortunately, research about the transport properties on S-doped FeSe is still left blank. In this paper, we report the detailed study of transport properties of FeSe$_{1-x}$S$_x$ ($x$ = 0, and 0.14) single crystals grown by vapor transport method. The S doping is found to suppress the structural transition from $\sim$ 86 K in FeSe to $\sim$ 49 K ($x$ = 0.14), with only slight enhancement of $T_c$. A  linear magnetoresistance is observed in both crystals, and is found to be triggered by the structural transition. The linear MR and related discussion indicate the possible existence of Dirac-cone like band structure with ultrahigh mobility, which may come from the ferro-orbital ordering induced band shift. The mobility of the Dirac-cone-like band is found to decrease with S doping. Besides, the influence of S doping to the transport properties and band structure of FeSe is also studied and discussed, and is attributed to the suppression of orbital ordering from the chemical pressure of S doping.

\section{experiment}
 Single crystals with nominal composition FeSe$_{1-x}$S$_x$ ($x$ = 0, and 0.2) were grown by the vapor transport method \cite{BöhmerPRB}. Fe and Se/S powers were thoroughly mixed by grounding in glove box for more than 30 mins, and sealed in an evacuated quartz tube together with mixture of AlCl$_3$ and KCl powders. The quartz tube with chemicals was loaded into a horizontal tube furnace with one end heated up to 400$^\circ$C, while the other end was kept at 250$^\circ$C. After more than 35 days, single crystals with dimensitions in millimeter can be obtained in the cold end. The actual composition of the single crystals was characterized by energy-dispersive x-ray spectroscopy (EDX). The actually S doping level, obtained by an average of several different points, is $\sim$ 0.14 for the nominal 20\% sulfur doped crystal.

 Structure of the crystals was characterized by means of X-ray diffraction (XRD) with Cu-K$\alpha$ radiation. Microstructural and compositional investigations of the crystals were performed using a scanning electron microscope (SEM) equipped with EDX. Magnetization measurements were performed using a commercial SQUID magnetometer (MPMS-XL5, Quantum Design). The Hall resistivity $\rho$$_{yx}$ and magnetoresistance $\rho$$_{xx}$ were measured by using the six-lead method with the applied field parallel to $c$-axis and perpendicular to the applied current. In order to decrease the contact resistance, we sputtered gold on the contact pads just after the cleavage, then gold wires were attached on the pads with silver paste, producing contacts with ultralow resistance ($<$100 $\mu\Omega$).

\begin{figure}\center
\includegraphics[width=8.5cm]{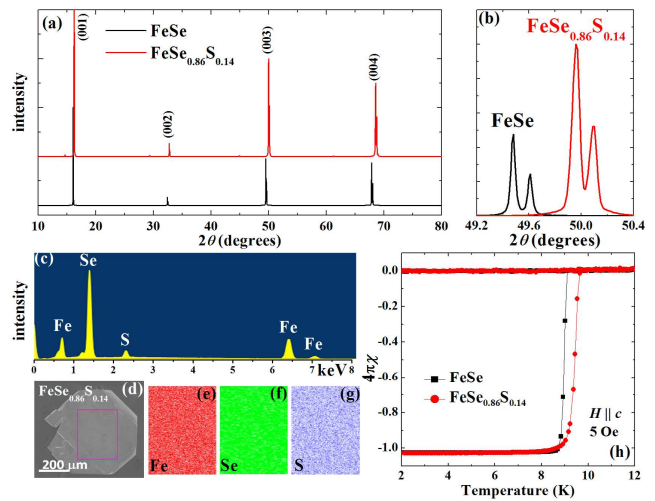}\\
\caption{(Color online) (a) Single crystal X-ray diffraction patterns of the FeSe and FeSe$_{0.86}$S$_{0.14}$ single crystals. (b) Enlarged part of (003) peaks. (c) EDX spectra and (d) SEM photo of FeSe$_{0.86}$S$_{0.14}$. Compositional mappings of (e) Fe, (f) Se, (g) S in the selected rectangular region in (d). (h) Temperature dependence of magnetic susceptibility $\chi$ (after considering the demagnetization effect) at 5 Oe for FeSe and FeSe$_{0.86}$S$_{0.14}$ single crystals.}\label{}
\end{figure}
\section{results and discussion}
Fig. 1(a) shows the single crystal XRD patterns for the FeSe and FeSe$_{0.86}$S$_{0.14}$ single crystals. Only the (00$l$) peaks are observed, suggesting that the crystallographic $c$-axis is perfectly perpendicular to the plane of the single crystals. With S doping, the positions of (00$l$) peaks obviously shift to higher values of 2$\theta$, which can be seen more clearly in the enlarged part of (003) peaks shown in Fig. 1(b). Fig. 1(d) shows the SEM photo of the FeSe$_{0.86}$S$_{0.14}$ single crystal. From the corresponding EDX spectra shown in Fig. 1(c), the spectrum peak from S can be clearly identified, which together with the obvious shift in XRD pattern prove that the S is successfully doped into the crystal. Compositional mappings in the selected rectangular region of Fig. 1(d) are shown in Figs. 1(e)-(g), which prove that Fe, Se and S are almost homogeneously distributed in the crystal.

Fig. 1(h) shows the temperature dependence of magnetic susceptibility $\chi$ at 5 Oe for FeSe and FeSe$_{0.86}$S$_{0.14}$ single crystals. FeSe displays a superconducting transition temperature $T_c$ $\sim$ 9.0 K, which is slightly enhanced after S doping to $\sim$ 9.5 K in FeSe$_{0.86}$S$_{0.14}$. Such a slight enhancement of $T_c$ is similar to the previous report \cite{AbdelPRB}. Taking the criteria of 10 and 90\% of the magnetization result at 2 K, the superconducting transition width, $\Delta$$T_c$, for both crystals is estimated $\leq$ 0.5 K, which manifests the single superconducting phase of our single crystals. More information about the magnetization properties of the FeSe single crystal was reported in our previous publication \cite{PhysRevB.92.144509}.
\begin{figure}\center
\includegraphics[width=8.5cm]{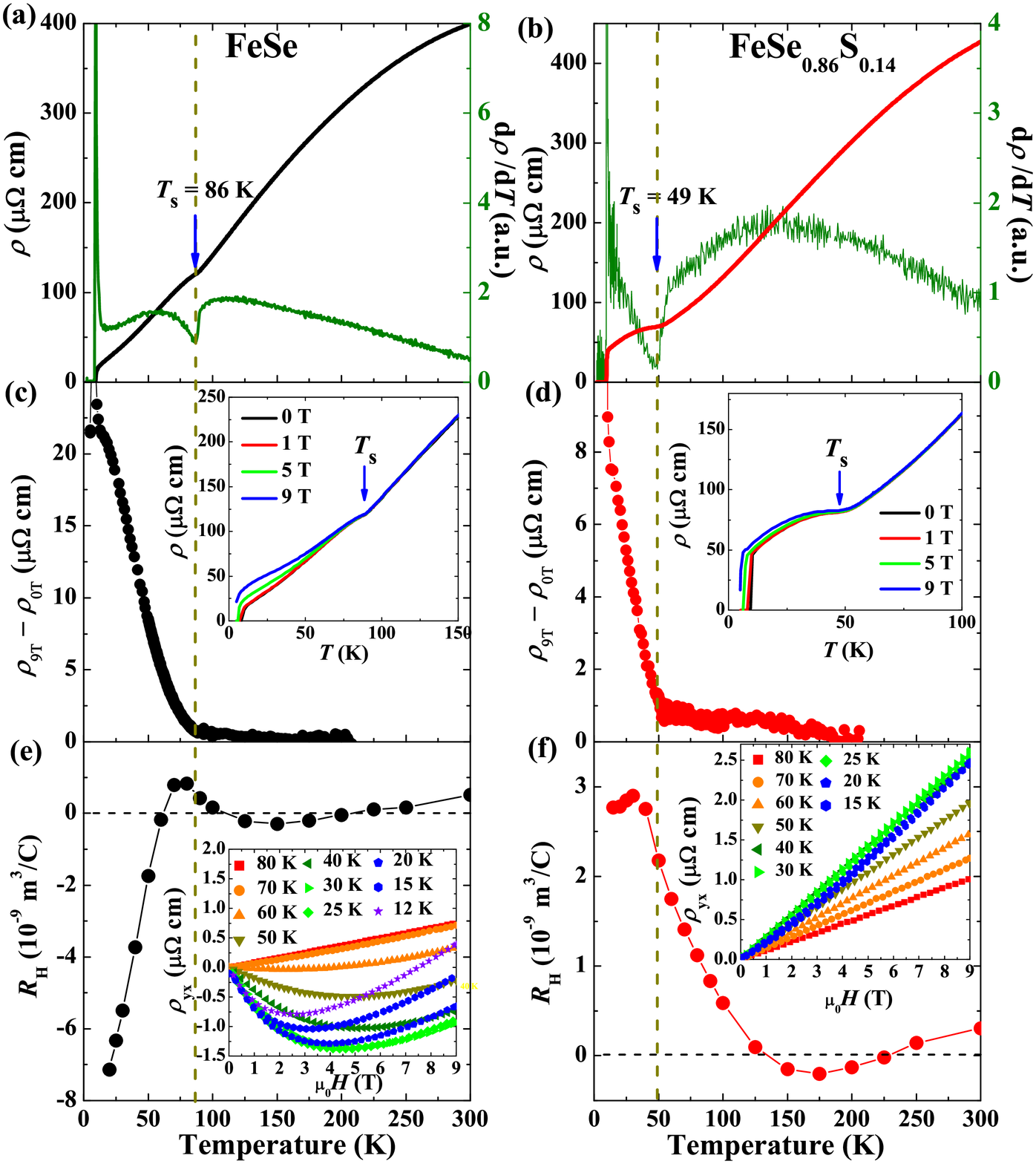}\\
\caption{(Color online) Temperature dependence of in-plane resistivity and its 1st derivative for (a) FeSe and (b) FeSe$_{0.86}$S$_{0.14}$ single crystals, respectively. Insets of (c) and (d) show the $\rho - T$  curves measured under 0, 1, 5, and 9 T magnetic field for FeSe, and FeSe$_{0.86}$S$_{0.14}$. Main panels of  (c) and (d) show the difference between $\rho(T)$ measured under 9 and 0 T. Insets of (e) and (f) show the Hall resistivity $\rho_{yx}$ at several temperatures for FeSe, and FeSe$_{0.86}$S$_{0.14}$, respectively. Main panels of (e) and (f) show corresponding Hall coefficients $R_H$ for both crystals.}\label{}
\end{figure}

Fig. 2(a) shows the temperature dependence of in-plane resistivity for FeSe single crystal, which manifests a metallic behavior until temperature decreasing to $T_c$. The residual resistivity ratio RRR, defined as $R$(300 K)/$R$(10 K), is of $\sim$ 33. Such a large value of  RRR is close to the recent report of a clean single crystal with impurities less than one per 2000 Fe atoms \cite{Kasahara18112014}, which manifests that our single crystal contains few impurities/defects. An obvious kink-like behavior related to the structural transition is also observed which is similar to previous reports \cite{Kasahara18112014}, and can be seen more clearly in the derivative of temperature dependent resistivity, d$\rho$/d$T$, plotted also in Fig. 2(a). The structural transition temperature, $T_s$, defined as the temperature at which d$\rho$/d$T$ takes its minimum value as masked by the arrow, is $\sim$ 86 K for FeSe. After S doping, $T_s$ is considerably suppressed to $\sim$ 49 K for FeSe$_{0.86}$S$_{0.14}$ as shown in Fig. 2(b), although the value of $T_c$ is affected little. According to the recent ARPES measurement, the suppression of $T_s$ may come from the suppression of orbital ordering by chemical pressure either from physical pressure or from S doping \cite{Watsonarxiv}.

To get more information about the difference before and after structural transition, we also performed temperature dependent resistivity measurements under magnetic fields up to 9 T for FeSe and FeSe$_{0.86}$S$_{0.14}$, as shown in the insets of Figs. 2(c) and (d), respectively. For both crystals, resistivities are almost field independent above $T_s$, and can fall into one curve. Such behavior is common in compensated metal since the amounts of electrons and holes are equal. Besides, the S is isovalent to Se, which introduces no extra electrons or holes. Thus, both crystals show field-independent resistivity above $T_s$. However, an obvious divergence for resistivity under different fields can be observed immediately below $T_s$ for both crystals. This behavior can be seen more clearly from the difference between $\rho$($T$) measured under a field of 9 T and 0 T as shown in the main panels of Figs. 2(c) and (d). Obviously, ($\rho_{9T}$ - $\rho_{0T}$) increases drastically below $T_s$. For FeSe, such a divergence under $T_s$ is explained by the electron-hole inequality proposed based on the asymmetry of d$I$/d$V$ curves in scanning tunneling spectroscopy (STS) \cite{PhysRevB.92.060505}. Our results manifest that such an asymmetry may be also present in FeSe$_{0.86}$S$_{0.14}$, and happens immediately below $T_s$ for both crystals, which indicates that such behavior is triggered by the structural transition.

In order to get more comprehensive understanding of the transport properties, we also measured the Hall resistivity, $\rho_{yx}$, for both crystals at temperatures above and below $T_s$. Typical results of field dependent $\rho_{yx}$ at low temperatures for FeSe and FeSe$_{0.86}$S$_{0.14}$ single crystals are shown in the insets of Figs. 2(e) and (f), respectively. Field dependence of  $\rho_{yx}$ for FeSe becomes obviously nonlinear below 80 K, and even changes to negative values at low temperatures, which is similar to previous reports \cite{HuynhPRB,WatsonPRL}. On the other hand, after S doping, $\rho_{yx}$ keeps positive and linearly increases with magnetic field at all temperatures for FeSe$_{0.86}$S$_{0.14}$. Hall coefficients $R_H$ can be simply obtained from $R_H$ = $\rho$$_{yx}$/$\mu_0H$, and are shown in the main panels of Fig. 2(e) and (f). For the nonlinear $\rho_{yx}$ at low temperatures in FeSe, $R_H$ is simply calculated from the linearly part at low fields. The values of $R_H$ are small above 100 K, and show similar temperature dependent behavior, changing sign twice, for both crystals. Such temperature dependent behavior of $R_H$ for FeSe is very similar to previous report  \cite{WatsonPRL}. The small absolute value of $R_H$ can be easily understood by considering a simply compensated two-band model containing equal numbers of electron- and hole-type charge carriers with similar mobility. And the sign change means that the temperature dependence of the mobilities for electrons and holes are not completely equal. $R_H$ of FeSe$_{0.86}$S$_{0.14}$ shows almost identical behavior as that of FeSe, confirming the isovalent nature of S-doping. It ensures that the effect of S doping is mainly caused by the chemical pressure rather than the charge carriers doping.

On the other hand, when temperature decreases between 100 K and $T_s$, $R_H$ for both crystals shows a positive value, and increases with decreasing temperatures. Such behavior may come from the change of mobility, which means that the hole-type carriers become more dominant at this temperature region. When temperature decreases below $T_s$, $R_H$ of FeSe shows a quick decrease, and changes sign from positive to negative, forming a hump-like structure in $R_H$ - $T$ curve. Such a hump-like structure is also observed in FeSe$_{0.86}$S$_{0.14}$, although the value of $R_H$ keeps positive at all temperatures. The decrease of $R_H$ (even a sign change in FeSe) indicates that the electron-type charge carriers contribute more (even becomes dominant in FeSe) to the transport properties before temperature decreasing to $T_c$. Here, we want to point out that the peak position of the hump in $R_H$ - $T$ is slightly below $T_s$, which is reasonable because the peak in $R_H$ should show up when the decreasing trend conquers the increasing trend of $R_H$ although the contribution of decreasing $R_H$ may begin from $T_s$. Such noticeable temperature dependent $R_H$ reflects the multiband nature of both crystals. Besides, recent mobility spectrum analysis and three-band model fitting on FeSe attributed the striking change in $R_H$ to the emergence of a small electron band with ultrahigh mobility \cite{HuynhPRB,WatsonPRL}. We will discuss this point in detail below in the part of MR.

\begin{figure}\center
\includegraphics[width=8.5cm]{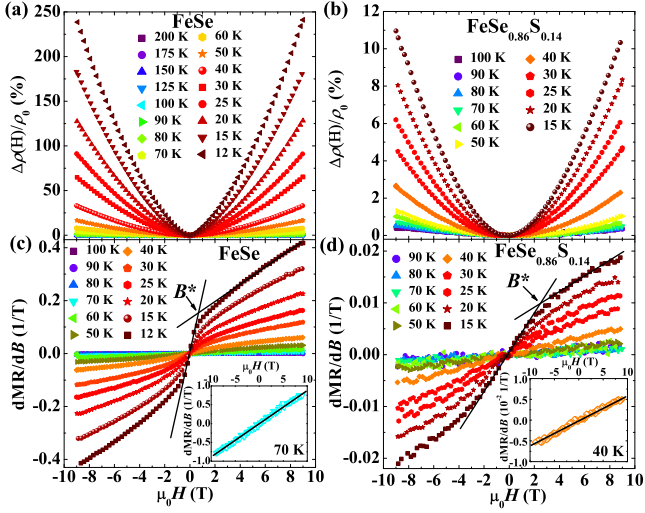}\\
\caption{(Color online) Magnetic field dependence of  magnetoresistance (MR = ($(\rho(H)-\rho(0))/\rho(0)$) for (a) FeSe,  (b) FeSe$_{0.86}$S$_{0.14}$, and field derivative MR (dMR/d$B$) for (c) FeSe,  (d) FeSe$_{0.86}$S$_{0.14}$ single crystals measured at different temperatures, respectively. The insets of (c) and (d) show the magnetic field dependent dMR/d$B$ measured at 70 K for FeSe, and 40 K for FeSe$_{0.86}$S$_{0.14}$, respectively.}\label{}
\end{figure}

To further investigate the influence of structural transition and S doping to the transport properties and band structure of FeSe, we also studied the MR of both crystals. Fig. 3(a) show the magnetic field dependence of  magnetoresistance (MR=($(\rho(H)-\rho(0))/\rho(0)$) for FeSe single crystal at different temperatures. Obviously, MR of FeSe at temperature higher than $T_s$ shows a relative smaller value, $\leq$ 1\%, which is expected for conventional compensated metal with modest mobilities. While, the value of MR increases dramatically with temperature below $T_s$, and reaches a large value over 200\% at 12 K under 9 T. The large value of MR at low temperatures is similar to previous reports \cite{HuynhPRB,WatsonPRL}. similar temperature dependent behavior of MR is also observed in FeSe$_{0.86}$S$_{0.14}$ as shown in Fig. 3(b) that MR shows a small value at temperatures above 40 K, and increases quickly below $T_s$. However, the value of MR for FeSe$_{0.86}$S$_{0.14}$ is more than 1 order smaller than that of FeSe.

\begin{figure}\center
\includegraphics[width=8.5cm]{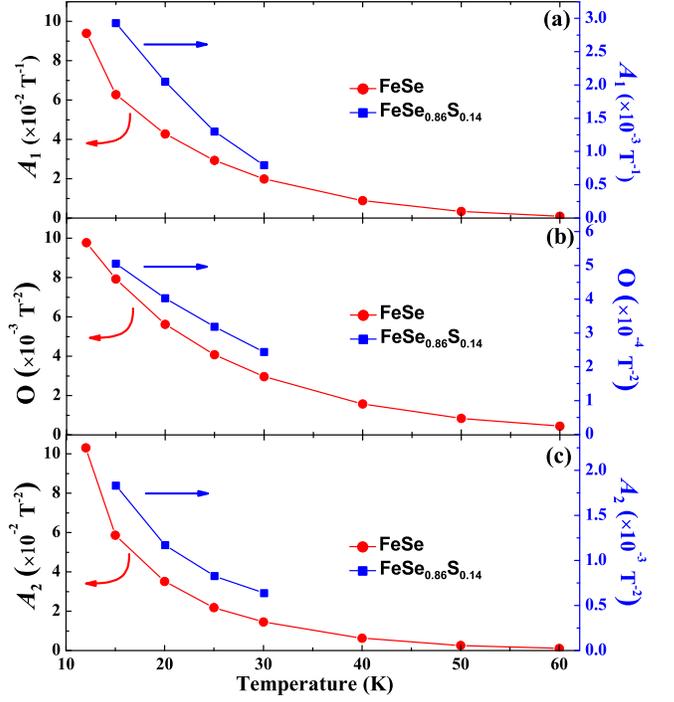}\\
\caption{(Color online) Coefficients (a)$A_1$ (b) $O$ and (c) $A_2$ for FeSe and FeSe$_{0.86}$S$_{0.14}$ obtained by fitting magnetic field dependent MR using Eq. (1). }\label{}
\end{figure}
More interestingly, the MR below $T_s$ tends to increase with magnetic field in a more linear relation at high field region for both crystals, whereas a small parabolic-like bend just remains at low fields. This is in sharp contrast to the semiclassical quadratic field dependence of MR, in which MR generally develops in proportion to $B^2$ over the entire field range. Such behavior can be witnessed more evidently in the first-order derivative d($MR$)/d$B$ as shown in Figs. 3(c) and (d) for FeSe and FeSe$_{0.86}$S$_{0.14}$, respectively. d($MR$)/d$B$ linearly increases with magnetic field at small fields, which indicates a classic $B^2$ dependence of MR. While, above a characteristic field $B^*$, d($MR$)/d$B$ saturates to a much reduced slope. Such an abrupt reduction of slope in MR with increasing field, masked by the solid lines in Fig. 3(c), is usually understood by the contribution of a linear field dependent MR plus a quadratic term. Therefore, the MR below and above $B^*$ can be usually expressed as \cite{KuoPRB}

\begin{equation}
\label{eq.1}
  MR = \begin{cases}
    A_2[\mu_0H]^2 & \mu_0H < B^*,\\
    A_1\mu_0H + O[\mu_0H]^2 & \mu_0H > B^*,
  \end{cases}
\end{equation}
where $A_2$ is the coefficient for $B^2$ terms when $\mu_0H < B^*$; $A_1$ and $O$ are the coefficients for $B$ linear and $B^2$ terms when $\mu_0H > B^*$. By using the formula above, we fitted the magnetic field dependent MR at different temperatures for both FeSe and FeSe$_{0.86}$S$_{0.14}$ single crystals. The obtained parameters are compared and shown in Fig. 4. It is obvious that all the parameters decrease with increasing temperature. The values of $O$ are smaller than those of $A_1$ for both crystals, which indicates that the linear MR is more dominant in the transport properties for $\mu_0H > B^*$. Besides, the values of $A_1$ and $O$ for FeSe$_{0.86}$S$_{0.14}$ are much smaller than those of FeSe, which manifests that the observed suppression of MR by S-doping has effects on both linear and the quadratic terms.

The linear MR observed in single crystal is usually interpreted by considering a quantum limit where all the carriers occupy only the lowest Laudau level (LL) \cite{AbrikosovPRB,AbrikosovEPL}. This situation usually happens when the field is very large and the difference between the zeroth and first Landau levels $\Delta_{LL}$ exceeds the Fermi energy $E_F$ and the thermal fluctuations $k_BT$. In such a quantum limit, MR can no longer be described in the framework of the conventional Born scattering approximation, and is instead expressed as:
\begin{equation}
\label{eq.2}
MR=\frac{1}{2\pi}(\frac{e^2}{\varepsilon_\infty\hbar v_F})^2\frac{N_i}{en^2}Bln(\varepsilon_\infty),
\end{equation}
where $N_i$ is the density of  scattering centers, $n$ is the carrier density, $v_F$ is the Fermi velocity and $\varepsilon_\infty$ is the high-frequency dielectric constant \cite{AbrikosovPRB,AbrikosovEPL}. In a conventional parabolic band, the LL is proportional to $B$, $\Delta_{LL}$ = e$\hbar B$/$m^*$, where $m^*$ is the effective mass. To satisfy the quantum limit, i.e., $\Delta_{LL}>k_BT$,  a very large value of magnetic field is needed. Thus, the linear MR coming from the quantum limit is difficult to be observed in a moderate field range. By contrast, the linear MR was identified in low field region in some materials hosting Dirac fermions with linear energy dispersion, such as graphene \cite{NovoselovNature}, topological insulators \cite{TaskinPRL}, Ag$_{2-\delta}$(Te/Se) \cite{XuNature}, $\alpha$-(BEDT-TTF)$_2$I$_3$ \cite{Kobayashijpsj}, some layered compounds with two-dimensional Fermi surface (like SrMnBi$_2$) \cite{ParkPRL,WangPRB} and iron-based Ba(Sr)Fe$_2$As$_2$ \cite{HuynhPRL,IshidaPRB122,ChongEPL}, La(Pr)FeAsO \cite{PallecchiPRB,BhoiPrFeAsO}, and a related compound FeTe$_{0.6}$Se$_{0.4}$ \cite{SunPRB}. For the Dirac state, $\Delta_{LL}$ is described as $\Delta_{LL}$ = $\pm$$v_F$$\sqrt{2e\hbar B}$, leading to a much larger LL splitting compared with the parabolic band. Consequently, the quantum limit  can be achieved in low field region \cite{AbrikosovPRB}.

Now, we discuss a little more about the behavior of MR above $B^*$ for materials holding Dirac fermions. For materials with single band, all the carriers occupy only the lowest Laudau level when the $\Delta_{LL}$ is opened under magnetic field. In such a case, the high-field MR will show only linear behavior, and the d($MR$)/d$B$ is almost field independent. For materials with multiband, when the carriers in one or several bands fall into Dirac cone state, those carriers become Dirac fermions with very high mobility. In this situation, if the mobility of other carriers from normal bands is much smaller than the mobility of Dirac fermions, the Dirac fermions will be dominant in the transport, and the contribution of normal state carriers becomes negligible. Thus, the high-field MR will also show only linear behavior, and the d($MR$)/d$B$ is again field independent. On the other hand, if the mobility of carriers from normal bands is not much smaller than that of the Dirac fermions, like the case of FeSe (Based on the three-band model fitting \cite{WatsonPRL}, the mobility of carriers from normal bands is $\sim$ 7,000 cm$^2$/Vs at 20 K, and that of the possible Dirac fermions is $\sim$ 25,000 cm$^2$/Vs.), the transport results will manifest combined properties of both the normal carriers and Dirac fermions. Thus, the high-field MR shows a reduced slope, which is combined with the linear and quadratic terms, rather than the simple linear behavior. Actually, such behavior of reduced slope in MR is also observed before in other compounds like Sr(Ca)MnBi$_2$ \cite{WangPRB,WangCaMnBi} and Ba(Sr)Fe$_2$As$_2$\cite{KuoPRB,IshidaPRB122,ChongEPL}.

Here, we should point out that the contribution of linear MR is only observed at temperatures below $T_s$. As shown in the insets of Figs. 3(c) and (d), the two slopes behavior in d($MR$)/d$B$ is suppressed and replaced by a unique slope when temperature increases to 70 K for FeSe, and 40 K for FeSe$_{0.86}$S$_{0.14}$, respectively. It indicates that the emergence of Dirac fermions is triggered by structural transition.

\begin{figure}\center
\includegraphics[width=8.5cm]{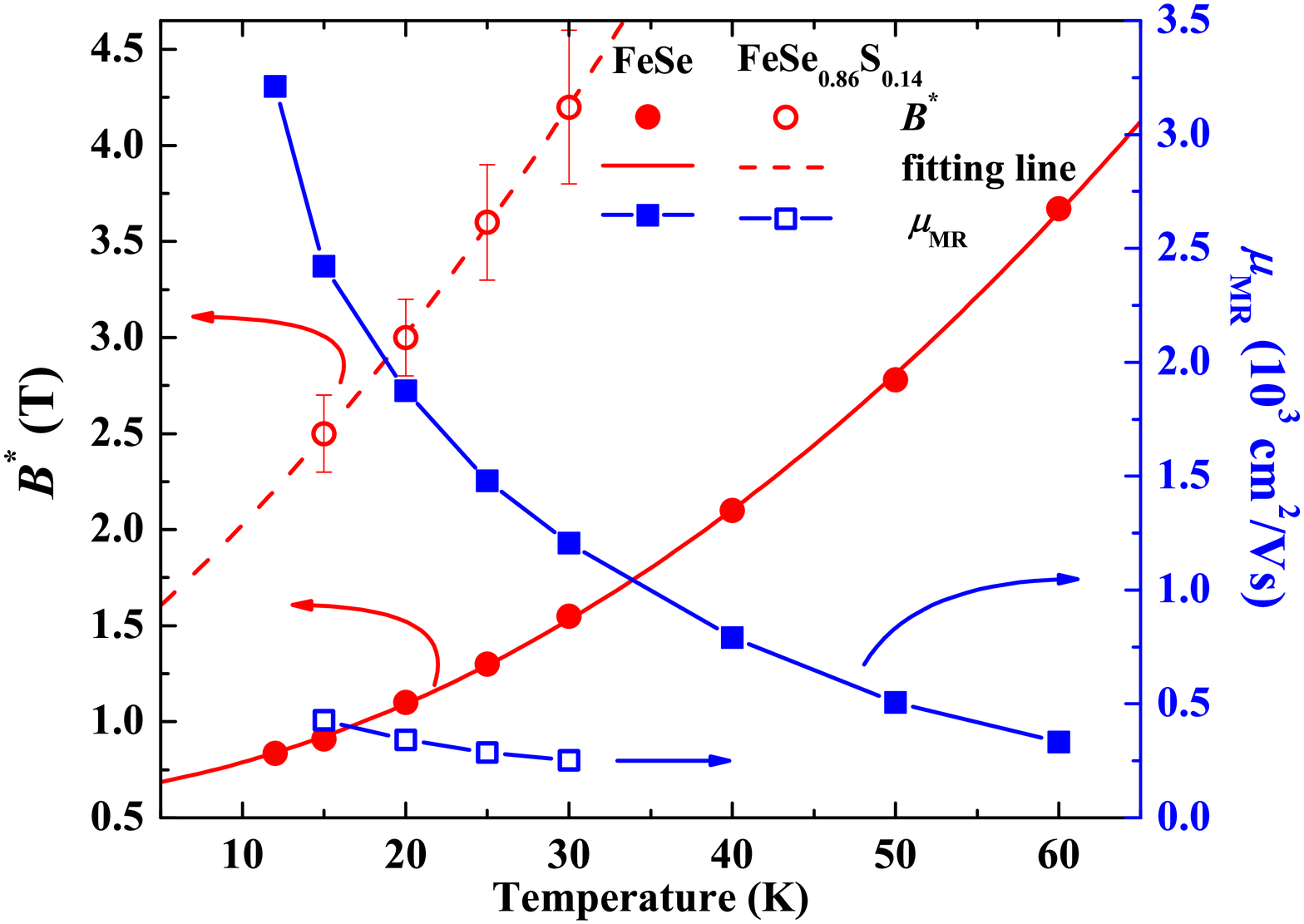}\\
\caption{(Color online) Temperature dependence of the characteristic field $B^*$ and the effective MR mobility $\mu_{MR}$ for FeSe (solid symbols) and FeSe$_{0.86}$S$_{0.14}$ (open symbols) single crystals, respectively. Red solid and dashed lines are the fitting of $B^*$ by $B^*$ = (1/2e$\hbar v^2_F$)($k_BT$+$E_F$)$^2$.}\label{}
\end{figure}
The temperature dependence of characteristic field $B^*$ for FeSe is shown in Fig. 5, which is obviously violating the linear relation expected from conventional parabolic bands, and can be well fitted by $B^*$ = (1/2e$\hbar v^2_F$)($k_BT$+$E_F$)$^2$ for the Dirac fermions as shown by the red solid curve in Fig. 5 \cite{TaskinPRL}. The good agreement of $B^*$ with the above equation supports the existence of possible Dirac fermions in FeSe. The fitting gives a large Fermi velocity $v_F$ $\sim$ 9.1 $\times$ 10$^4$ ms$^{-1}$, which is close to the previous reports in iron-based BaFe$_2$As$_2$ ($v_F$ $\sim$ 1.9 $\times$ 10$^5$ ms$^{-1}$) \cite{HuynhPRL}, SrFe$_2$As$_2$ ($v_F$ $\sim$ 3.1 $\times$ 10$^5$ ms$^{-1}$) \cite{ChongEPL}, and the related compound FeTe$_{0.6}$Se$_{0.4}$ ($v_F$ $\sim$ 1.1 $\times$ 10$^5$ ms$^{-1}$) \cite{SunPRB}. Actually, the emergence of an extra band with ultrahigh mobility after structural transition in FeSe is also supported by the mobility spectrum analysis and three-band model fitting \cite{HuynhPRB,WatsonPRL}. On the other hand, in a multiband system with both Dirac and coventional parabolic band where Dirac carriers are dominant in transport, the prefactor $A_2$ for the $B^2$ term is related to the effective MR mobility $\sqrt{A_2}$ = $\frac{\sqrt{\sigma_e\sigma_h}}{\sigma_e+\sigma_h}$($\mu_e$+$\mu_h$) = $\mu_{MR}$ \cite{TaskinPRL,KuoPRB}. The effective MR is smaller than the average mobility of carriers $\mu_{avg}$ = ($\mu_e+\mu_h$)/2, and gives an estimation of the lower bound. Temperature dependence of $\mu_{MR}$ was calculated and shown by the solid square in Fig. 5. The values of $\mu_{MR}$ mainly reside in the order of 10$^3$ cm$^2$/Vs similar to the estimation by mobility spectrum and three-band model fitting \cite{HuynhPRB,WatsonPRL}. $\mu_{MR}$ decreases with increasing temperature since thermal fluctuations smear out the LL splitting.

The temperature dependence of $B^*$ and $\mu_{MR}$ for FeSe$_{0.86}$S$_{0.14}$ are plotted in open symbols, and compared with those from FeSe shown in Fig. 5. Obvious, the values of $B^*$ is largely enhanced after S doping, and the fitting gives a Fermi velocity $v_F$ $\sim$ 7.4 $\times$ 10$^4$ ms$^{-1}$. The similar temperature dependence of $B^*$ in FeSe$_{0.86}$S$_{0.14}$ indicates that the Dirac fermions may also exist in this material. The reduced $v_F$ indicates that the Dirac-cone-like band structure is suppressed by S doping. Besides, $\mu_{MR}$ is also found to be reduced to a value smaller than 500 cm$^2$/Vs.

The Dirac cone state is also observed in iron pnictide BaFe$_2$As$_2$, and is found to be a consequence of the nodes in the SDW gap by complex zone folding \cite{RichardPRL}, and it can coexist with superconductivity in Ru-doped BaFe$_2$As$_2$ until the complete suppression of SDW \cite{TanabePRBCoexis,TanabePRBSuppre}. Similar results have also been reported in Ru-doped LaFeAsO \cite{PallecchiPRB}. Such a mechanism is not suitable for FeSe since no magnetic order occurs belows $T_s$ \cite{McQueenPRL}. The possible Dirac-cone-like band structure found here may come from the band shift, which is caused by ferro-orbital ordering. In detail, the d$_{yz}$ band in the nematic state shifts up around the $M_x$ point, while, the d$_{xz}$ band shifts downwards around the $M_y$ point and opens a hybridization gap with the d$_{xy}$ band, which enlarges the electron pocket at the $M_y$ point \cite{Zhangarvix}. In addition, the band structure calculation in orbital ordered state considering the orbital-spin interplay shows that the electron Fermi surface at $X$-point will be deformed to two pockets with Dirac-cone dispersions \cite{Onariarxiv}. It can also explain the observed suppression of Dirac-cone-like band structure by S doping that the band shift is reduced because of the suppression of orbital ordering by the chemical pressure from S doping \cite{Watsonarxiv}.

\begin{figure}\center
\includegraphics[width=8.5cm]{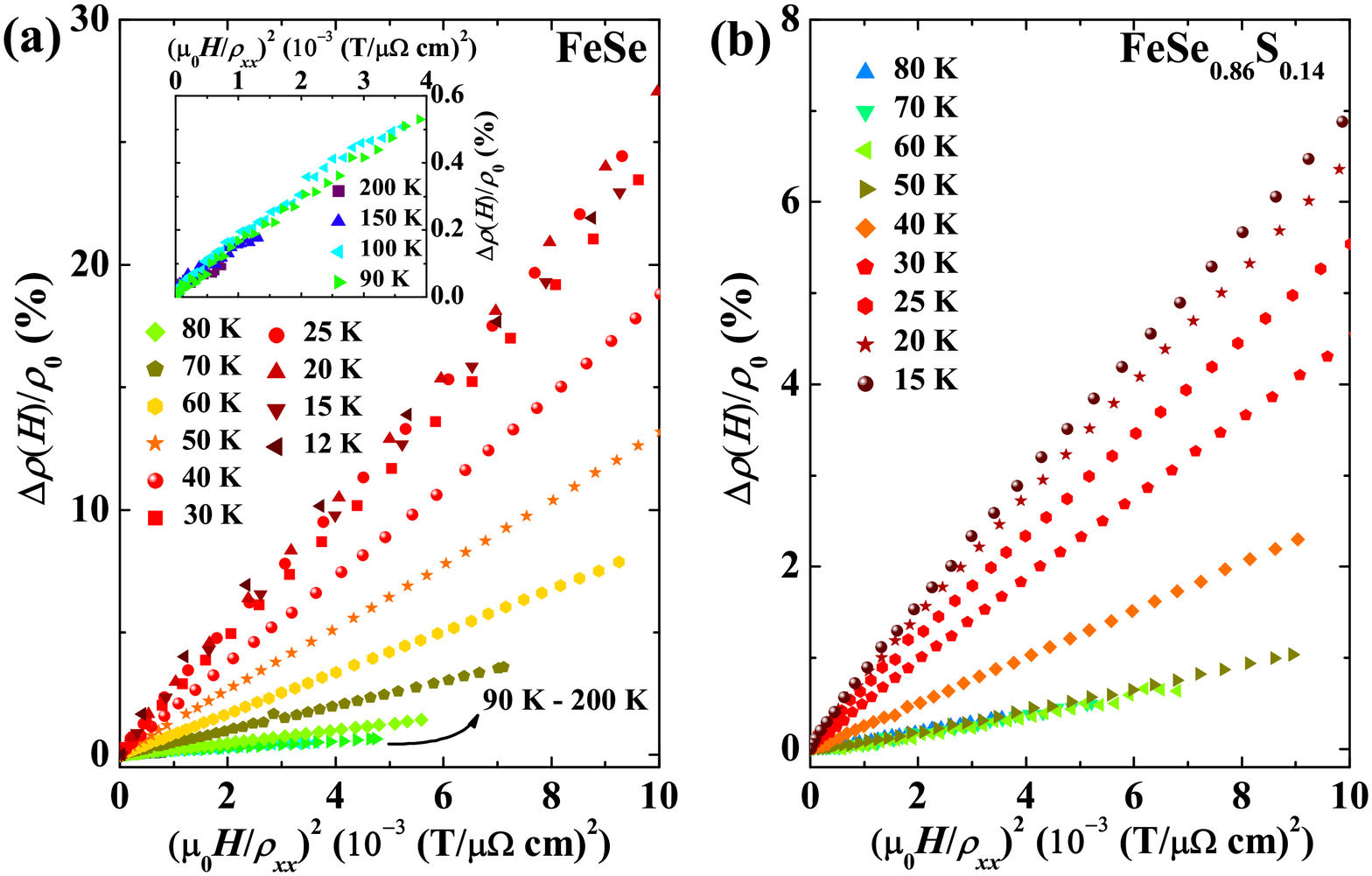}\\
\caption{(Color online) MR of (a) FeSe, and (b) FeSe$_{0.86}$S$_{0.14}$ single crystals plotted as a function of $(\mu_0H/\rho_{xx})^2$. Inset of (a) is the enlarged plot for the MR of FeSe measured at temperature ranging from 90 K to 200 K.}\label{}
\end{figure}
In principle, the band reconstruction induced by the orbital-ordering is the origin of dramatic change in transport properties when decreasing temperature below $T_s$. To get more comprehensive understanding of the S-doping effect, we replot the field dependent MR measured under different temperatures in Kohler's law \cite{ZimanBook}, in which the MR can be successfully scaled by $\Delta\rho(H)/\rho(0)=F(\omega_c\tau)=F[(\mu_0H/\rho(0))^2]$, where $F$ is a function of the cyclotron frequency $\omega_c$ and scattering time $\tau$ if the scattering rate for charge carriers are equal at all points on the Fermi surface. The Kohler's scaling of the MR for FeSe and FeSe$_{0.86}$S$_{0.14}$ are shown in Figs. 6(a) and (b), respectively. Since the existence of linear MR from Dirac-cone-like band structure will dramatically enhance the value of MR below $T_s$ especially under large fields, we mainly focus on the part below $B^*$.  Obviously, the MR for both crystals above $T_s$ ($T \geq $ 90 K for FeSe, $T \geq $ 50 K for FeSe$_{0.86}$S$_{0.14}$) can be well scaled into one curve, which means that the scattering rate is isotropic. For FeSe, the scale of MR by Kohler's law above $T_s$ can be seen more clearly in the enlarged plot as shown in the inset. Below $T_s$, the invalidity of Kohler's law can be observed for both crystals, indicating the anisotropic scattering rates. Such behavior is also observed in cuprates and iron pnicitides, and is usually attributed to the existence of hot and cold spots on the Fermi surface with anisotropic scattering rate \cite{KontaniRPP,KemperPRB}. The hot and cold spots are usually caused by the scattering from spin fluctuations, which is only observed below $T_s$ in FeSe \cite{BaekNatMater,BöhmerPRL}. The invalidity of Kohler's law in FeSe$_{0.86}$S$_{0.14}$ is also observed below $T_s$, which seems to suggest that the emergence of spin fluctuations is triggered by or strongly related to the structural transition/orbital ordering as also proposed by the NMR result in FeSe \cite{BöhmerPRL}.

Surprisingly, however, the Kohler's scaling becomes valid again when temperature decreases below $\sim$ 30 K in FeSe, and such behavior is absent in FeSe$_{0.86}$S$_{0.14}$. Previous report attributed the re-establishment of Kohler's law to the opening of a gap $\sim$ 8 meV above $T_c$ presumably at the hot spots of the Fermi surface based on their STM spectroscopy results \cite{PhysRevB.92.060505}. However, such large gap is not observed by other measurements or the other STM reports \cite{SongScience,SongPRL}. The re-establishment of Kohler's scaling can be also explained by the orbital-ordering-driven band reconstruction, in which the elliptical Fermi surface emerged below $T_s$ as explained above, will go on shrinking with decreasing temperature and eventually turns into two small circular ones as shown in Fig. 2(c) of ref. \cite{Zhangarvix}. Because of that, the scattering rate may become isotropic again because of the disappear of hot and cold spots. Furthermore, the orbital ordering was found to be suppressed after S doping based on the discussion above, which may not be strong enough to totally split the shrunk Fermi surface into two separated ones. Thus, the anisotropic scattering rate persists and the Kohler's scaling remains invalid at low temperatures below $T_s$ in FeSe$_{0.86}$S$_{0.14}$. Future experiments such as the ARPES or STM on S-doped FeSe are required to clarify this observation.

\section{conclusions}
In summary, we investigated the transport properties of FeSe$_{1-x}$S$_x$ ($x$ = 0, and 0.14) single crystals grown by vapor transport method. The structural transition was found suppressed from 86 K to 49 K after S doping, while the $T_c$ is only slightly enhanced from 9 K to 9.5 K. The non-linear field dependence of Hall resistivity observed below $T_s$ in FeSe was found to be replaced by linear behavior at all temperatures in FeSe$_{0.86}$S$_{0.14}$. A  linear MR triggered by the structural transition is observed in both FeSe and FeSe$_{0.86}$S$_{0.14}$ single crystals, indicating the possible existence of Dirac-cone state, which may come from the ferro-orbital ordering caused by band shift. The mobility of the Dirac-cone-like band was obviously suppressed by S doping. Besides, the invalid Kohler's scaling of MR was found at temperature below $T_s$ in both crystals. The Kohler's scaling becomes valid again at temperatures below 30 K in FeSe. However, such a re-establishment is absent after S doping. The observation and related discussion support that the orbital ordering motivated the band reconstruction in FeSe, and also manifest that the orbital ordering in FeSe can be suppressed by chemical pressure, e.g. S doping or physical pressure, which is promising for future understanding of the origin of the nematic state in FeSe as well as its relation to the novel superconductivity.

\acknowledgements
Y.S. gratefully appreciates the support from Japan Society for the Promotion of Science.

\bibliography{references}

\end{document}